\documentclass[oldversion, letter, longauth]{aa} 
\usepackage{graphicx}
\usepackage{txfonts}
\begin{document}
   \title{Uncovering the kiloparsec-scale stellar ring of NGC5128\thanks{Based on observations made with ESO Telescopes at the La Silla or Paranal Observatories under programme IDs 066.C-0310 and 081.C-0477}}



   \author{J. T. Kainulainen \inst{1 \and 2} 
          \and
	  J. F. Alves \inst{3}
          \and 
          Y. Beletsky \inst{4}
          \and 
          J. Ascenso \inst{5}
          \and 
          J. M. Kainulainen \inst{6}
          \and 
          A. Amorim \inst{7}
          \and 
          J. Lima \inst{7}
          \and 
          R. Marques \inst{8}
          \and 
          J. Pinh\~ao \inst{8}
          \and 
          J. Rebord\~ao \inst{9}
          \and 
          F. D. Santos \inst{7}
          }

   \offprints{J. Kainulainen}

   \institute{Observatory, P.O. Box 14, FIN-00014 Univ. of Helsinki, Finland \\
              \email{jtkainul@mpia-hd.mpg.de}
         \and TKK/Mets\"ahovi Radio Observatory, Mets\"ahovintie 114, FIN-02540 Kylm\"al\"a, Finland
         \and Calar Alto Observatory, Centro Astron\'omico Hispano, Alem\'an, C/q Jes\'us Durb\'an Rem\'on 2-2, 04004 Almeria, Spain
         \and European Southern Observatory (ESO), Alonso de Cordova 3107, Santiago, Chile
         \and Harvard-Smithsonian Center for Astrophysics, 60 Garden Street, Cambridge, MA 02138, USA
         \and TKK/Department of Radio Science and Engineering, P.O. Box 3000, FIN-02015 TKK, Finland
         \and SIM-IDL, Faculdade de Ci\^encias da Universidade de Lisboa, Ed. C8. Campo Grande 1749-016 Lisbon, Portugal
         \and LIP-Coimbra, Department of Physics, University of Coimbra, 3004-516 Coimbra, Portugal
         \and INETI, Estrada da Portela, Zambujal-Alfragide, Apartado 7586, 2720-866 Amadora, Portugal 
  }	
   \date{Received <>; accepted <>}
  \abstract{We reveal the stellar light emerging from the kiloparsec-scale, ring-like structure of the NGC5128 (Centaurus A) galaxy in unprecedented detail. We use arcsecond-scale resolution near infrared images to create a ``dust-free" view of the central region of the galaxy, which we then use to quantify the shape of the revealed structure. At the resolution of the data, the structure contains several hundreds of discreet, point-like or slightly elongated sources. The typical extinction-corrected surface brightness of the structure is $K_\mathrm{S} \approx 16.5$ mag/arcsec$^2$, and we estimate the total near infrared luminosity of the structure to be $M \approx -21$ mag. We use diffraction limited ($FWHM$ resolution of $\approx 0.1^{\prime\prime}$, or 1.6 pc) near infrared data taken with the NACO instrument on the VLT to show that the structure decomposes into thousands of separate, mostly point-like sources. According to the tentative photometry, the most luminous sources have $M_\mathrm{K} \approx -12$ mag, making them red supergiants or relatively low-mass star clusters. We also discuss the large-scale geometry implied by the reddening signatures of dust in our near infrared images.
}  	
   \maketitle

%
%

\section{Introduction} 

%
The Centaurus A galaxy (NGC5128, Cen A hereafter) is one of the most widely-recognized extragalactic objects in the sky. Being the most nearby giant elliptical containing an active galactic nucleus (AGN) and an associated jet, it has been the target of a wide range of AGN studies in the past (see Israel \cite{1998A&ARv...8..237I} for a review). On a large scale, Cen A features a giant elliptical galaxy that has a several kiloparsec wide disk of gas and dust in its central region. The disk is suggested to be a remnant of a merger of a small, gas-rich galaxy and a giant elliptical (Baade \& Minkowski \cite{1954ApJ...119..215B}) some $2-7 \times 10^8$ years ago (e.g. Quinn \cite{1984ApJ...279..596Q}; Quillen et al. \cite{1993ApJ...412..550Q}; Sparke \cite{1996ApJ...473..810S}). The merger is generally believed to be responsible for the AGN activity in {Cen A}. 

The central disk of gas and dust in Cen A is manifested by an opaque and rather chaotic dust absorption lane at optical wavelengths (see Marconi et al. \cite{2000ApJ...528..276M} for high resolution optical images). Because of the obscuration, the detailed stellar content and thereby the role of the merger for the star formation in Cen A has remained quite elusive. Observations of dust and molecular line emission have revealed a kiloparsec-scale, S- or bar-like structure that extends to about one kiloparsec radius, coinciding with the dust lane (e.g. Mirabel et al. \cite{1999A&A...341..667M}; Leeuw et al. \cite{2002ApJ...565..131L}; Espada et al. \cite{2009ApJ...695..116E}). The structure is known to exhibit star formation, as evidenced e.g. by Pa$\alpha$ emission (Marconi et al. \cite{2000ApJ...528..276M}). The velocity field of the gas in the structure has been quite successfully modeled as an inner part of a rotating, warped disk viewed from a high inclination angle (Bland et al. \cite{1987MNRAS.228..595B}; Nicholson et al. \cite{1992ApJ...387..503N}; Quillen et al. \cite{1992ApJ...391..121Q}). The warped disk model can also qualitatively describe the obscuration at visual wavelengths (Quillen et al. \cite{1993ApJ...412..550Q}). Most recently, Quillen et al. (\cite{2006ApJ...645.1092Q}, Q06 hereafter) used Spitzer mid-infrared images to show that the dust emission originates primarily from a parallelogram-shaped structure, and that there is a void in the dust distribution between $6 - 50^{\prime\prime}$ ($100 - 800$ pc). 


In this Letter, we report near infrared observations revealing the stellar light from the kiloparsec-scale, ring-like structure in the central region of Cen A. These observations unveil the geometry of the structure in unprecedented detail and for the first time using ground-based data. We describe the geometry of the observed structure in the context of the warped disk model developed for Cen A by Quillen et al. (\cite{1993ApJ...412..550Q}). Furthermore, we present tentative, adaptive optics assisted near infrared data confirming the stellar nature of the sources in the structure. In \S 2 of this Letter we describe the observations, and in \S 3 we present the results and discuss them briefly. Throughout the paper, we adopt the distance of 3.42 Mpc to Cen A (Ferrarese et al. \cite{2007ApJ...654..186F}). 

\section{Observations}      
\label{sec:observations}

\subsection{NTT/SOFI observations}  

\begin{figure*}
   \centering
\includegraphics[width=1.063\columnwidth]{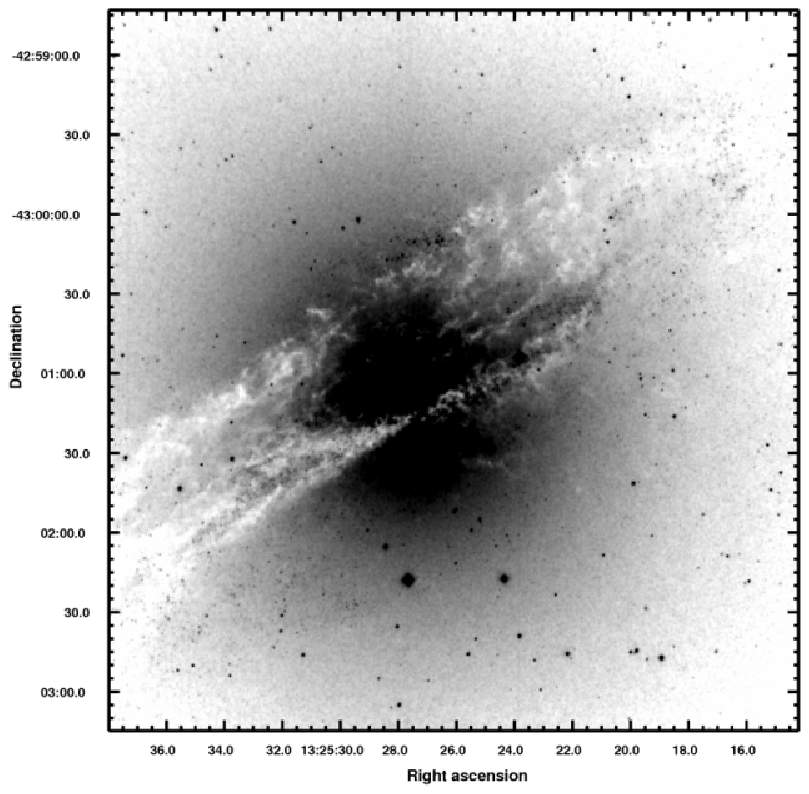}
\includegraphics[width=0.937\columnwidth]{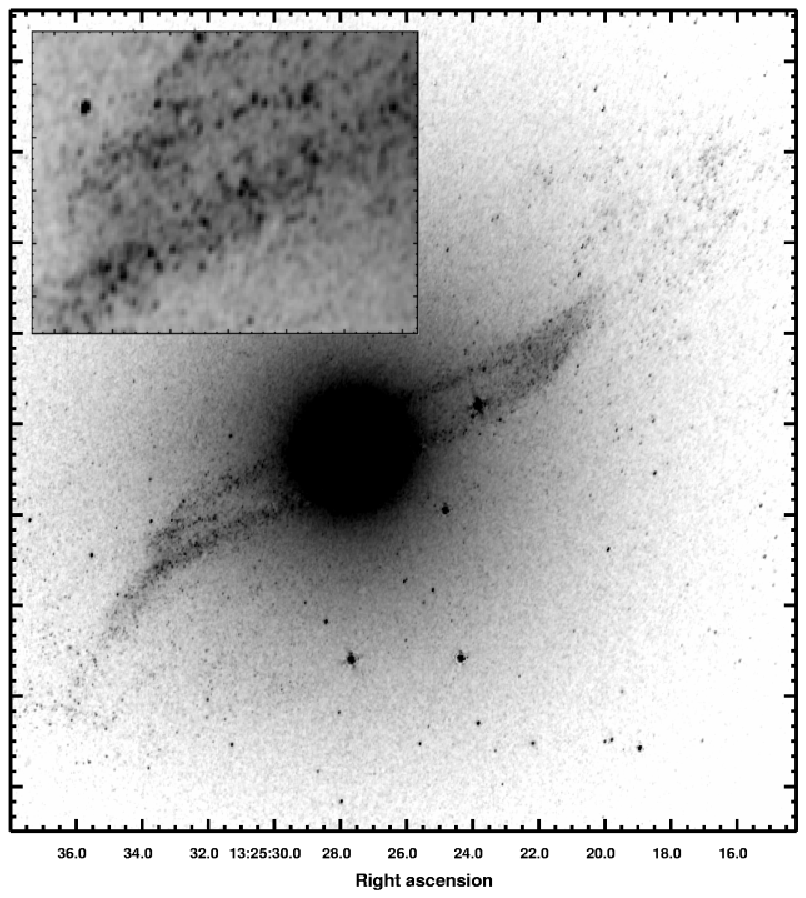}
      \caption{The observed SOFI data and the revealed ring-like stellar structure of the Cen A galaxy. The intensity scale is inverted for visibility. {\bf Left: } The observed $J$ band image. {\bf Right: } The same after the first-order extinction correction in which $E_\mathrm{J-H}$ image has been used as a template for dust (see \S\ref{subsec:dereddening} in text). The inset shows a blowup of the SE part of the structure.             }
         \label{fig:fig1}
   \end{figure*}

We use the photometrically calibrated $JHK_\mathrm{S}$ band surface brightness images, presented in Beletsky et al. (\cite{beletsky09}), as the starting point for our analysis. These images, taken with the SOFI instrument on 3.5-m New Technology Telescope during $5.-8.$ March 2003, cover $\sim 4' \times 4'$ region from the central part of Cen A. For the details of the data and data reduction we refer to Beletsky et al. \cite{beletsky09}). The $J$ band image is shown in Fig. \ref{fig:fig1}.

\subsection{VLT/NACO observations}  

We obtained $J$ ($1.265\pm 0.25$ $\mu$m), $H$ ($1.66\pm 0.33$ $\mu$m), and $K_\mathrm{S}$ ($2.18\pm 0.35$ $\mu$m) band, high-resolution images of the Cen A galaxy on June 30th 2008, using the adaptive-optics assisted imager NACO at the Nasmyth-B focus of the ESO/VLT 8-m telescope, ``Yepun",  located at Cerro Paranal, Chile. The NACO (NAOS-CONICA) instrument is equipped with an adaptive optics system, NAOS (Rousset et al. \cite{2003SPIE.4839..140R}), which provides both visible and infrared wavefront sensing and illuminates the CONICA camera (Lenzen et al. \cite{2003SPIE.4841..944L}; Hartung et al. \cite{2003SPIE.4841..425H}) equipped with an Aladdin $1024\times1024$ pixel InSb array detector. The images were collected with the L27 camera ($28^{\prime\prime} \times 28^{\prime\prime}$, 27.12 mas/pix), which provided fully sampled images according to the Nyquist sampling criterion. The bright star at the SE corner of the field served as the natural guide star. A full description of the data acquisition and detailed analysis will appear in Ascenso et al. (in prep.).

\section{Results and discussion}      

\subsection{Kiloparsec-scale ring of Cen A in near infrared}  
\label{subsec:dereddening}


We subtracted the observed SOFI surface brightness images from each other, yielding $J-H$ and $H-K$ color maps. We processed the color maps further by calculating the color excess of each image pixel with respect to the color of the elliptical galaxy in that pixel:
\begin{equation}
E_\mathrm{i-j} = (m_\mathrm{i}- m_\mathrm{j})_\mathrm{observed} - (m_\mathrm{i}- m_\mathrm{j})_\mathrm{elliptical}.
\label{eq:E}
\end{equation}
To estimate the color of the elliptical galaxy, we calculated the average observed color as a function of radius inside spherical annuli of $2^{\prime\prime}$, centered at the galaxy nucleus. In the region where the dust lane prevents the determination of the color profile, we evaluated the profile from the fit of a second order polynomial to the color profile outside the dust lane. 
The color excess maps derived in this way are morphologically similar to the color excess maps shown in Beletsky et al. (\cite{beletsky09}), except for the large scale gradient due to the elliptical, which is eliminated from our maps. The images are included in the online version of this Letter (Fig. 4).


It is possible to make a first-order absorption correction to the observed surface brightness images using the color excess maps as templates for dust. If the absorbing dust is located in front of all stellar emission, the extinction in band $i$ is related to the color-excess $E_\mathrm{i-j}$ via equation:
\begin{equation}
A_\mathrm{i} = X_\mathrm{ij} \times E_\mathrm{i-j},
\label{eq:color-excess}
\end{equation}
where $A_\mathrm{i}$ is extinction in band $i$ and $X_\mathrm{ij} = (1+\tau_\mathrm{j}/\tau_\mathrm{i})^{-1}$. The constant $X_\mathrm{ij}$ can be calculated by adopting the extinction law e.g. by Rieke \& Lebofsky (\cite{1985ApJ...288..618R}) which yields $X_\mathrm{JH} \approx X_\mathrm{HK} = 2.6$. We note that Eq. (\ref{eq:color-excess}) is not generally valid for dust clouds embedded in galaxies, because it is usually unrealistic to assume that the clouds are located in front of the stellar emission of the galaxy. However, we argue that in the exceptional case of Cen A this simplification is, in fact, not unrealistic. We justify the argument and discuss it in Section \ref{subsubsec:dustgeometry}.


We used the color excess maps to perform an extinction correction for the observed images, as facilitated by Eq. (\ref{eq:color-excess}). To illustrate the result, the extinction corrected $J$ band image is shown in Fig. \ref{fig:fig1}. These images reveal a bright, ring-like structure coinciding with the southern part of the dust lane. The structure is clearly similar to the parallelogram-shaped feature previously detected in the Spitzer data at $3.6-8 $ $\mu$m (see Q06 for the $8$ $\mu$m image). Especially, there is an obvious similarity between the extinction corrected images and the Spitzer 3.6 $\mu$m image that is dominated by emission of the same stellar component as the NIR data. However, the sub-arcsecond resolution SOFI data (Fig. \ref{fig:fig1}) reveals the structure in unprecedentedly high detail and completely undisturbed by thermal dust emission. The structure contains several hundreds of sources that are point-like or slightly elongated at the resolution of the SOFI data. The nature of these sources is further discussed in Section \ref{subsec:results-naco}. 


We compared the integrated intensity in the observed and extinction corrected images to examine the total attenuation caused by the dust lane. This yielded attenuations of 0.19 mag, 0.26 mag, and 0.4 mag in $K_\mathrm{S}$, $H$, and $J$, respectively. We also estimated the total NIR luminosity of the ring-like structure. To do this, we subtracted the intensity profile of the elliptical galaxy from the extinction corrected images using a simple spherical model. Fig. \ref{fig:par} shows the resulting image in $K_\mathrm{S}$ band. In addition to the ring structure, the image shows numerous additional sources, especially in the NW part of the image. Typical intensities in the most prominent regions are $K_\mathrm{S} \approx 16.5$ mag / arcsec$^2$. Integration of the intensity over the ring structure yields total NIR luminosities of $M_\mathrm{J} = -20.5$ mag, $M_\mathrm{H} = -21.6$ mag, and $M_\mathrm{K} = -21.8$ mag. We note that these estimates are subject to several sources of uncertainty, especially including the extinction correction. Therefore, they should be considered merely as first-order estimates.

\subsection{Geometry inferred by the data}  
\label{subsec:geometry}


The warped disk model developed for gas and dust in Cen A is described by a series of thin, co-centered rings whose inclination and position angles vary as a function of radius (Quillen et al. \cite{1992ApJ...391..121Q}; see also Q06). These rings define the central plane of the disk. When viewed from a high inclination angle, the changing sign of the disk inclination angle with respect to the line of sight causes twists in the disk, both in north and south sides of the nucleus. The dust present in the disk can qualitatively explain the geometry of the observed absorption and emission features (we refer to Q06 and references therein for a detailed description of the model). In the following, we examine our NIR data in the context of the warped disk model. 

\subsubsection{Shape of the ring-like structure}


The sub-arcsecond resolution view of the uncovered structure provides an excellent basis for quantifying its shape. Except for the ``twisted" SE and NW edges, the shape of the structure is quite well described by a ring with major and minor axes of $a = 67^{\prime\prime}$ and $b = 8.5^{\prime\prime}$ (1100 pc and 140 pc), respectively, and a diameter of $d = 4^{\prime\prime}$ (70 pc). The position angle of the ring is $25.5^\circ$, and in order to best match it with the observed structure, an offset of $1.5^{\prime\prime}$ north along the minor axis is required. The region inside which there are no sources has a major axis of $a \approx 50^{\prime\prime}$ ($\approx 800$ pc), in unison with Q06.

\begin{figure}
   \centering
   \includegraphics[width=\columnwidth]{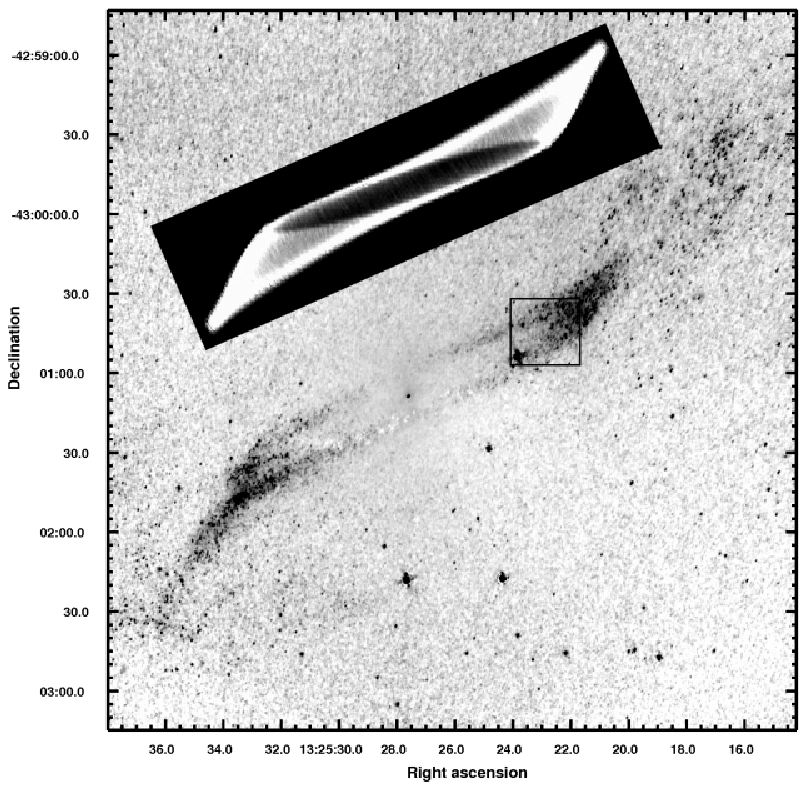}
   \includegraphics[width=0.65\columnwidth]{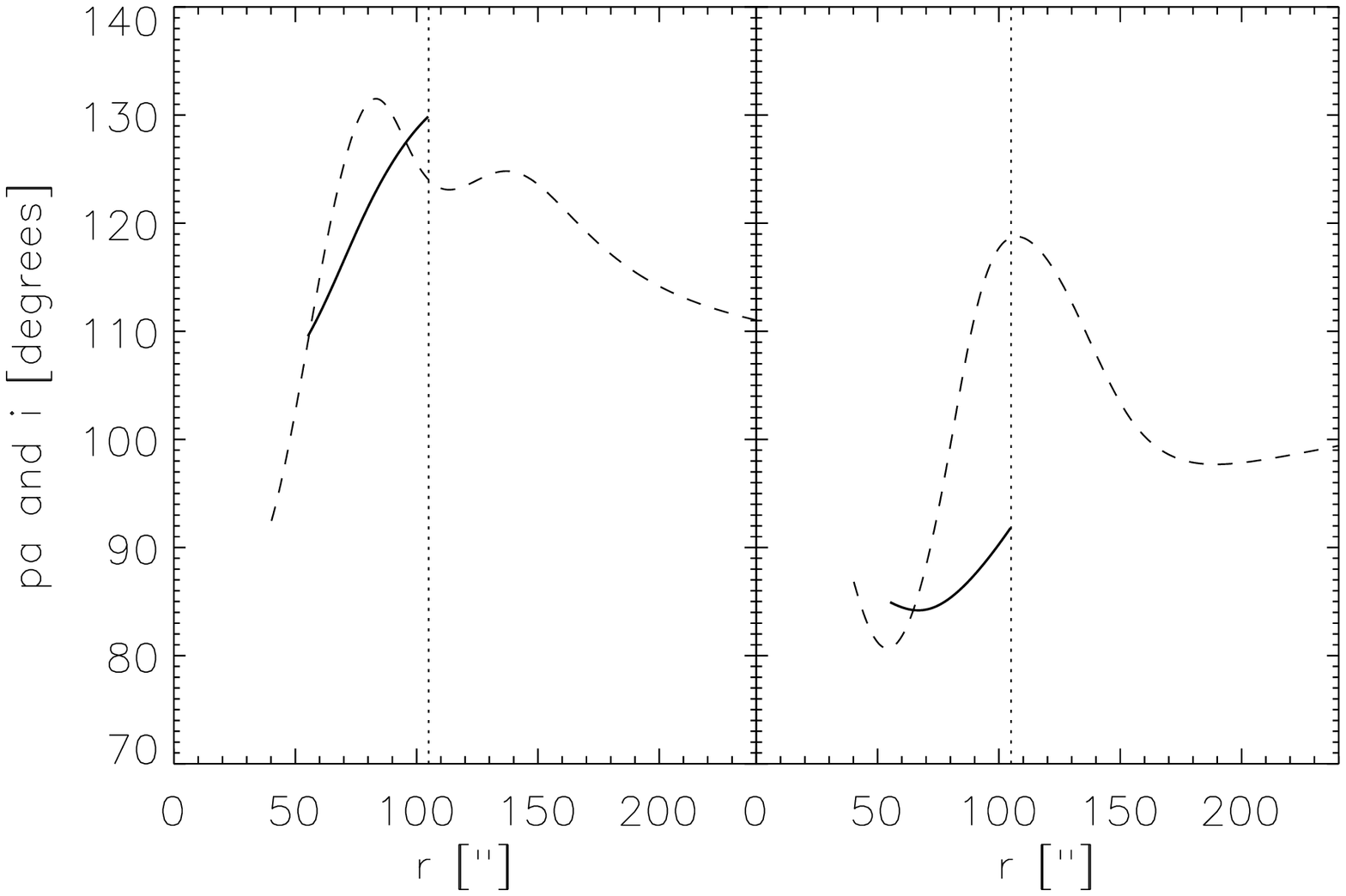}
      \caption{{\bf Top: }The extinction corrected $K_\mathrm{S}$ band image from which the intensity of the elliptical has been subtracted. The inset shows the fit to the image of a model composed of co-centric rings with varying inclination and position angles (see text). The small box shows the position of the NACO observations. {\bf Bottom: }The position and inclination angles of the best-fitting model for the ring structure. The dashed lines show the angles given by Q06.} 
      \label{fig:par}
\end{figure}

We used the ellipsoid subtracted images to determine the inclination and position angles, $\alpha(r)$ and $\omega(r)$, that define the warped disk model. We constructed the warped disk model inside $r<100^{\prime\prime}$ as explained in Q06 with the difference that in our model the emission from the disk was constant as a function of radius from the nucleus. In Q06, the angles $\alpha(r)$ and $\omega(r)$ were defined by giving a set of points $[\alpha, r]$ and $[\omega, r]$ and by making an interpolation between the points using a spline function. We construct our model by minimizing the $\chi^2$ value between the observed and model images where $\alpha(r)$ and $\omega(r)$ are free parameters. Fig. \ref{fig:par} shows the best-fitting model and the angles $\alpha(r)$ and $\omega(r)$ resulting from the fit. We note that the fit is not satisfactory in describing the detailed intensity structure of the ring, but the model reproduces the general shape of the ring seemingly well.

\subsubsection{Dust geometry}
\label{subsubsec:dustgeometry}

The location of the dust clouds responsible for the absorption features in Cen A can be examined in a model-dependent manner by assuming that the dust is located in the warped disk as described in Q06. For each line-of-sight in the observed images, it is straightforward to calculate the position, or the range of positions, where that particular line-of-sight is connected to the warped disk structure. For example, it is evident that within the warped disk framework, the dust features north of the nucleus are associated with the fold at $r \approx 80-150^{\prime\prime}$.

It is interesting to consider the fraction of stellar light emitted in front of the dust structures at each position of the observed images. To estimate the fraction, a radial model for the stellar emission is needed. We adopted the radial luminosity profile given by Marconi et al. (\cite{2006A&A...448..921M}) who fitted the intensity profile using an oblate spheroidal function. Using this model, we integrated the emission of the radial emission model along each line-of-sight towards the dust features to the point closest to the observer where that line-of-sight connects with the warped disk structure. 
The fractions of foreground emission resulting from this calculation are between $1-20$ \%. In particular, the fraction of the foreground light in the fold at the northern side of the nucleus is less than 10 \%. Typical attenuation caused by the clouds in the field is less than a magnitude, i.e. less than $\sim 60$ \% in flux. Therefore, the intensity observed in any on-cloud pixel is unlikely to be dominated by the flux of the foreground stars.

  \begin{figure}
   \centering
   \includegraphics[width=\columnwidth]{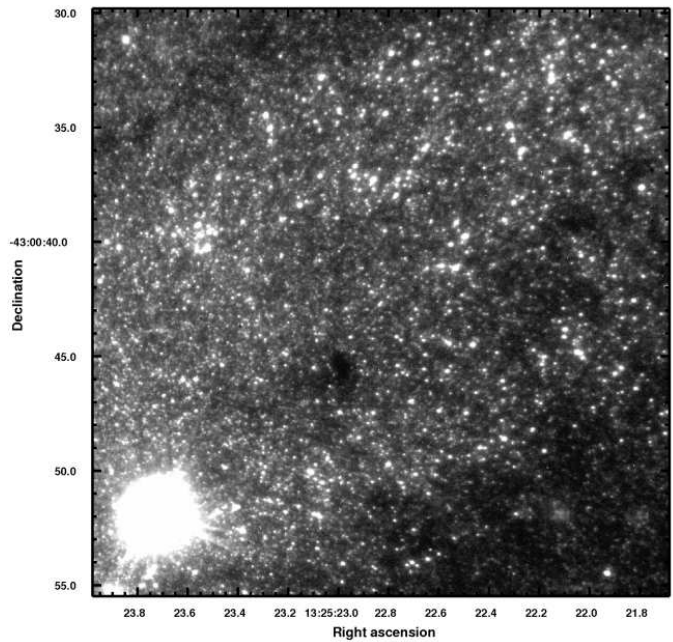}
      \caption{$K_\mathrm{S}$ band, adaptive optics assisted image of the NW end of the ring-like structure taken with the NACO instrument on the VLT (see Fig. \ref{fig:par} for the frame position). The brightest sources in the image are $K_\mathrm{S} \approx 16$ mag (discluding the saturated star in the lower left corner). Most sources have PSFs consistent with point sources.}
         \label{fig:naco}
   \end{figure}

While the discussion above relies on a certain model for the dust geometry, it is possible to use the wavelength dependency of the observed attenuation to make a model independent, first-order approximation of the overall dust geometry (e.g. Gallagher \& Hunter \cite{1981AJ.....86.1312G}; Howk \& Savage \cite{1997AJ....114.2463H}; Kainulainen et al. \cite{2008A&A...482..229K}). To examine the wavelength dependency of the reddening, we measured the ratio of color excesses, $E_\mathrm{J-H} / E_\mathrm{H-K}$, at each pixel in the fold north of the nucleus where $E_\mathrm{H-K} > 0.2$. This resulted in a mean ratio of $E_\mathrm{J-H} / E_\mathrm{H-K} = 1.7$, a value similar to that implied by the standard Galactic reddening laws, e.g. 1.7 by Rieke \& Lebofsky (\cite{1985ApJ...288..618R}). If the dust extinction law in Cen A is similar to the Galactic one, such a high ratio is expected only if the fraction of the foreground light is low. Thus, the color excess ratio we calculated for the dust in Cen A supports a geometry where the dust is dominantly in front of most stellar emission, and thereby the warped disk model. Furthermore, the low fraction of foreground light suggests that it may not be unreasonable to assume that the color excess is a linear measure of the dust column density in the particular case of the Cen A dust lane.

\subsection{Tentative results from high resolution NIR data} 
\label{subsec:results-naco}


While a more complete analysis of our NACO data will be presented in a forthcoming paper (Ascenso et al., in prep.), we use the data in this Letter to show that the ring harbors a large number of sources that are point-like at parsec-scale resolution. Fig. \ref{fig:naco} shows the observed NACO $K_\mathrm{S}$ band image (see Fig. \ref{fig:naco} for the position of the frame). The image is filled with apparently point-like sources, distributed throughout the observed field. However, the brightest sources are clearly arranged to follow the ring structure seen in SOFI images. Comparison of the image to the SOFI data shows how the sources in SOFI images decompose into several or even tens of individual sources. In addition to the stellar sources, the image contains extended obscuration features due to dust. The features closely follow the morphology of the color excess maps made from SOFI data, although the superior sensitivity of the NACO data makes the features more prominent. Our tentative PSF-photometry of the $K_\mathrm{S}$ band image contains $\sim 3000$ sources above the 5-$\sigma$ detection limit. The brightest sources are $m_\mathrm{K} \approx 16$ mag while the majority is $m_\mathrm{K} \approx 18.5$ mag. Given that the SOFI data suggests extinction of $A_\mathrm{K} \approx 0.3$ mag in the least obscured part of the field, the absolute magnitudes of the brightest objects are $M_\mathrm{K} \approx -12$ mag. According to the tentative photometry, almost all sources have PSFs consistent with point sources. These properties suggest that the brightest sources are red supergiants, or rather low-mass star clusters, located within the ring. 

In conclusion, the observations presented in this Letter provide a new view of the structure and the stellar content of the Cen A galaxy. We have uncovered and quantified the shape of a large-scale stellar structure in the interior of Cen A, previously exposed only by space-based mid-infrared data. Such information can be used to better constrain the dynamical models and star-forming history of the Cen A galaxy.


\begin{figure*}
   \centering
   \includegraphics[width=0.992\columnwidth]{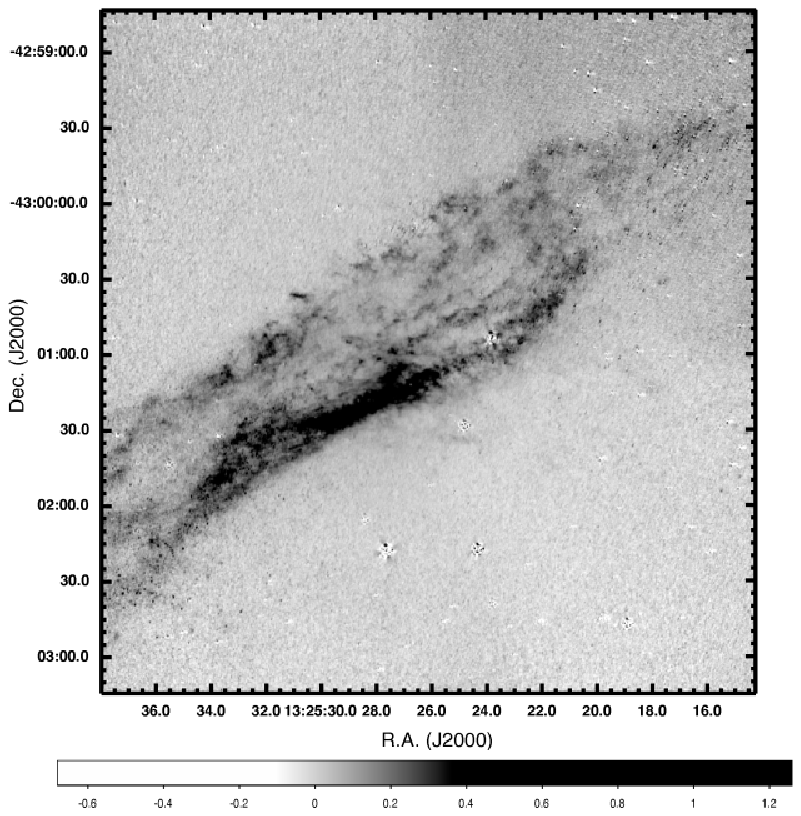}
\includegraphics[width=1.008\columnwidth]{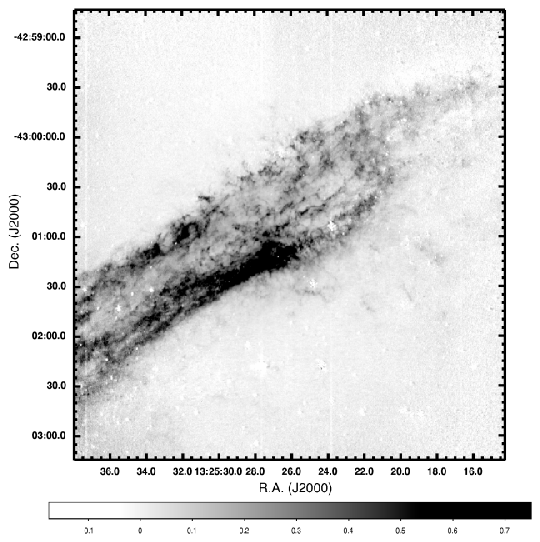}
      \caption{Near infrared color excess maps of the Centaurus A galaxy (see text for the derivation). {\bf Left: }$E_\mathrm{H-K}$. {\bf Right: }$E_\mathrm{J-H}$.} 
      \label{fig:color-excess}
\end{figure*}

\begin{acknowledgements}

The authors would like to thank A. Quillen who kindly provided the electronic Spitzer and HST images to be compared with our data. 

\end{acknowledgements}


\end{document}